\documentstyle[eqsecnum,preprint,aps,prd,epsfig]{revtex}
\preprint{DAMTP-R98/24, DTP 98-31}
\draft
\tighten
\begin{document}
\def\sqr#1#2{{\vcenter{\hrule height.3pt
      \hbox{\vrule width.3pt height#2pt  \kern#1pt
         \vrule width.3pt}  \hrule height.3pt}}}
\def\square{\mathchoice{\sqr67\,}{\sqr67\,}\sqr{3}{3.5}\sqr{3}{3.5}}
\def\today{\ifcase\month\or
  January\or February\or March\or April\or May\or June\or July\or
  August\or September\or October\or November\or December\fi
  \space\number\day, \number\year}

\def\Bbb{\bf}


\title{Superconducting $p$-branes and extremal black holes}

\author{A. Chamblin{$^{1}$}, R. Emparan{$^2$} and G.W. Gibbons{$^1$}}

\address {\qquad \\ {$^1$} DAMTP, Silver Street\\
Cambridge, CB3 9EW, England  
\qquad \\ {$^2$} Department of Mathematical Sciences\\
University of Durham\\
Durham DH1 3LE, England
}

\maketitle

\begin{abstract}

In Einstein-Maxwell theory, 
magnetic flux lines are `expelled' from a black hole
as extremality is approached, in the sense
that the component of the field strength normal to the horizon goes to zero. 
Thus, extremal black holes are found to 
exhibit the sort of `Meissner effect' which is 
characteristic of superconducting media.  We review some of the 
evidence for this
effect, and do present new evidence
for it using recently found black hole solutions in string theory 
and Kaluza-Klein
theory.  We also present some new solutions, which arise naturally in
string theory, which are {\it non}-superconducting extremal black holes.
We present a nice geometrical interpretation of these effects derived by 
looking
carefully at the higher dimensional configurations from 
which the lower dimensional
black hole solutions are obtained.  We show that other extremal  
solitonic objects in string theory (such as $p$-branes) can
also display  superconducting properties.  In particular, we argue
that the relativistic London equation will hold on the worldvolume of
`light' superconducting $p$-branes (which are embedded in flat space), 
and that minimally coupled zero modes
will propagate in the adS factor of the near-horizon geometries of
`heavy', or gravitating, superconducting $p$-branes.

\end{abstract}


\section{Introduction}

As is well known, the phenomenon known as `superconductivity' was first 
discovered (and named) in 1911 by H.\ Kammerlingh-Onnes.  
Kammerlingh-Onnes, in the course of studying the electric resistance of certain 
metals which were cooled to liquid helium
temperatures, found that the resistance of mercury dropped drastically as 
the temperature was reduced from $4^{\circ}K$ to $3^{\circ}K$.  Later authors
found that the temperature range over which the drop in resistivity occurs is 
extremely small.  Thus, people were led to discover the first well-understood
property of superconducting media:  Below a certain critical temperature 
($T_c$),
the electric resistance of the medium is zero (to within experimentally 
relevant
bounds).  This behaviour is of course the origin of the term, superconductor.

On the other hand, given a superconducting medium at some temperature 
$T<T_c$,
it is always possible to get rid of the superconductivity by applying a 
minimum
magnetic field $B>B_c$, where $B_c(T)$ is some critical value of the magnetic
field which depends on the temperature $T$.  The destruction of 
superconductivity
by a sufficiently strong magnetic field, together with the fact that the 
superconductor has zero resistance, leads one inevitably to the conclusion 
that the magnetic induction must vanish inside a superconductor, i.e., 
$B=0$.  This property of superconductors, which is actually experimentally
observed (i.e., a magnet will `float' above a superconducting medium), is 
known
as the `Meissner effect'.  The Meissner effect is succinctly expressed by the 
statement
that a superconductor displays perfect diamagnetism.  It is this property of
superconducting media which is the principal focus of this paper.  In fact,
in this paper we shall use the terms `perfect diamagnet' and `superconductor'
interchangeably, even though technically perfect conductivity is only a 
necessary (not sufficient) condition for perfect diamagnetism.  

One may view superconductivity at 
various levels. One may begin by constructing a 
purely phemonological macroscopic
theory in which Maxwell's 
equations are taken as fundamental and one supplements
them with constitutive relations, of which the
the most useful is the London equation. One may then
pass to a classical thermodynamic formulation
of the phenomenon. Finally one may attempt
to identify the  quantum mechanical microscopic 
degrees  of freedom responsible. In this paper
we shall mainly be concerned with the phemonological
theory. We will establish the existence in classical
supergravity theories an analogue of the usual Meissner effect.
We will also have some suggestions as to how 
the purely phenomenological theory may be extended
to a thermodynamic and quantum mechanical theory.

In fact the behaviour of magnetic field lines in the presence of strong
gravitational fields has been investigated for some time by many authors (see, 
e.g., \cite{wald,klk,ernst,ew,bicak}).
In particular, in 1974 Wald \cite{wald} studied the behaviour of Maxwell test
fields in the presence of a rotating black hole described by the Kerr 
solution.
Using the fact that a Killing vector in a vacuum spacetime acts as a 
vector potential for the Maxwell test field, it is not hard to see that as 
the
hole is `spun up' and approaches extremality, the component of the magnetic
field $B$ normal to the horizon tends to zero; thus, the flux lines are 
expelled
in the extremal limit and the hole behaves like a perfect diamagnet.

This effect was noticed and then confirmed in Einstein-Maxwell theory, to 
linear order in the magnetic field, by Bi\v{c}{\'a}k and Dvo\v{r}{\'a}k 
\cite{bicak}. In particular, they studied Reissner-Nordstrom holes
in the presence of magnetic fields induced by current loops. In \cite{bicak} 
very nice pictures are presented for the field lines around
a hole as it approaches extremality, so the emergence of the Meissner
effect can actually be seen.  More recently, the authors of \cite{cham}
considered an Abelian Higgs vortex in the Reissner-Nordstrom background.
It was shown that in the extreme limit (but not near extremality)
{\it all} of the fields associated with the vortex 
(both the magnetic and scalar degrees of freedom) are expelled from 
the horizon of the black hole. The magnetic and scalar fields always 
`wrap around' the horizon in the extremal limit.

In this paper we shall first review the evidence that (light) $p$-branes are 
superconducting (Section \ref{light}), and then attempt to extend the analysis 
to include the effect of self-gravitation (Section \ref{selfg}). The appearance 
of a form of the Meissner effect on the extremal horizon of a brane (Section 
\ref{strmeiss}) leads us to perform a comprehensive analysis of magnetic fields 
in the vicinity of extremal horizons (Section \ref{extrmeiss}).
We establish the existence of this effect
in widely generic settings, which include Kaluza-Klein and string theories.  
Moreover,
we also present some exact solutions for extremal black holes in external 
fields which exhibit this Meissner effect. These should serve to dispel the 
notion that the effect is an artifact of the linearized approximation to the 
theory which could disappear after including the backreaction. We also address 
(Appendix \ref{nonsuc}) some subtle examples where apparently the field 
expulsion breaks down. A closer examination shows, however, that in those 
examples one should not have expected the expulsion to happen in the first 
place, because of an interaction induced by the presence of a 
Chern-Simons term.

\section{Superconducting extended objects: Light branes}\label{light}

We begin with a description of the superconducting properties
of {\it light} branes.  That is, in this section we ignore the coupling 
of the $p$-branes to gravity, so that we may think of the branes
as extended, sheet like objects (of zero thickness) 
moving in a flat spacetime background, with dynamics described by a
Dirac-Born-Infeld action.  In the next section, we will consider the
superconducting properties of spacetimes describing gravitating branes.
The superconducting properties of light branes have been 
discussed previously by Nielsen and Olesen \cite{nielsen,nielole}
and by Balachandran et al.\ \cite{bal} (superconducting vortices with non-zero 
thickness, such as those examined in \cite{witten}, will not be discussed 
here). Before reformulating their ideas in  a geometrical language 
which generalizes to the case of heavy branes we recall
for the readers' convenience some basic facts about the Meissner effect.

Phenomenological accounts of superconductivity
distinguish carefully between {\it perfect conductivity} , i.e. 
$\sigma \rightarrow \infty \Leftrightarrow {\bf E} = {\bf j} /\sigma =0$
and {\it perfect diamagnetism}. i.e. $\mu \rightarrow \infty \Rightarrow {\bf 
B}=0$. 
The former merely implies that ${\partial {\bf B} / \partial t}=0$
which in turn implies that an arbitrary amount of flux may be frozen 
into the  sample depending upon initial conditions. The latter however 
goes some way to implying the Meissner 
effect, i.e. that flux is expelled from the material so the 
superconducting state is independent of initial conditions.

One may  regard the Meissner effect as 
as a consequence of the so-called Becker-Heller-Sauter equation
\begin{eqnarray}\label{bhs}
{\bf E}= \lambda ^2 \,{\partial {\bf j} \over \partial t}
\end{eqnarray}
for some constant $\lambda$.  This yields (on use of charge conservation)
the freezing of magnetic flux:
\begin{eqnarray}\label{bhs2}
{\partial \over \partial t} \Bigl ({\bf B} + \lambda ^2 {\rm curl} \,{\bf j} 
\Bigr )=0.
\end{eqnarray}
The strictly stronger non-relativistic London equation
\begin{eqnarray}\label{lon}
\lambda ^2 {\rm curl } \,{\bf j} + {\bf B}=0
\end{eqnarray}
implies the Meissner effect more directly and yields, on use
of Faraday`s law ${\rm curl} ~ {\bf E} = - {\partial {\bf B} / \partial t}$
\begin{eqnarray}\label{lon2}
{\rm curl} \Bigl ( {\bf E} - \lambda ^2 {\bf j} \Bigr )=0 
\Rightarrow {\bf E} - \lambda ^2 {\bf j} = - {\rm grad }\psi
\end{eqnarray}
for some scalar field $\psi$. 

In a relativistic  generalization of the 
London equation is 
\begin{eqnarray}\label{rlon}
-{ 1 \over \lambda ^2} F_{\mu \nu} = \partial _\mu J_\nu - \partial _\nu J _ 
\mu
\end{eqnarray}
or
\begin{equation}\label{rlon2}
J_\mu = -{ 1\over \lambda ^2} A _\mu + {\partial}_{\mu} \Lambda
\end{equation}
for some function $\Lambda$.  Because ${\nabla}_{\mu}F_{\mu \nu} = -J_{\nu}$,
we have
\begin{equation}\label{rmeiss}
-{\nabla}^{2}J - \frac{1}{{\lambda}^{2}} = 0
\end{equation}
so the mass of the vector field is given as 
$\frac{1}{{\lambda}^{2}}$.
If $\Lambda =0$ and in the absence of charges
eqn.~(\ref{rlon}) is equivalent to eqn.~(\ref{bhs}) and eqn.~(\ref{lon}). In 
what follows we shall 
adopt eqn
(\ref{rlon}) as our criterion for superconductivity.

Balachandran et al.\ \cite{bal} have argued that eqn.~(\ref{rlon}) typically 
holds on the 
worldvolume $\Sigma$ of extended objects and Nielsen has shown, in the 
context
of Kaluza-Klein theory, that the relativistic London equation will hold on the
worldvolume of extended objects carrying Kaluza-Klein currents \cite{nielsen}.
The basic idea behind Nielsen's observation is that if 
$K^a$ is a Killing vector field generating a circle subgroup
of the Kaluza--Klein group $G$ of isometries of
a higher dimensional Kaluza-Klein manifold
${\cal E}$ and $\pi: {\cal E}  \rightarrow {\cal M}$ the projection onto
the spacetime  manifold  $\cal M$ and
\begin{equation}\
F^{ab} = \nabla ^a K^b - \nabla ^b K ^a
\label{Killer}\end{equation}
then
$\pi _\star F^{ab}$ is the Kaluza-Klein field strength on spacetime
$\cal M$. Now if $x: \Sigma \rightarrow {\cal E}$ is an immersion or 
embedding
of a $p+1$ dimensional submanifold or brane $\Sigma$ and $x_\pi= \pi \circ x$
the projection down to spacetime $\cal M$ then the pull back
$J= x^\star K$ of the Killing vector field $K$ to the worldvolume
$\Sigma$ yields (via Noether's theorem and  the field
equations for the embedding $x$) a conserved current $J$
on the worldvolume. But clearly pulling back eqn.~(\ref{Killer}) to the 
worldvolume
shows that $\pi _\star F$ and $J$ satisfy the London equation on $\Sigma$,
i.e. $\Sigma$ is superconducting with respect to the
the Kaluza-Klein current. We shall refer to this type
of superconductivity as Nielsen superconductivity.

So far we have not used any field equations, either for
the brane or for the background in which it moves. For light branes
in some fixed background the equations of motion of a brane
with vanishing Born-Infeld field on the worldvolume
and vanishing Ramond-Ramond fields in the bulk require that it 
be a minimal submanifold, a particular case of which is a totally
geodesic submanifold. In the next section we shall see
that some self-gravitating branes satisfying the Einstein equations
may be identifed 
with totally geodesic submanifolds. We can then see to what extent
they exhibit Nielsen superconductivity.

\section{Superconducting self-gravitating extended objects}\label{selfg}

In the last section we investigated the superconducting aspects of
extended objects which have decoupled from gravity.  This limit, where
the branes are `light' so that one may focus strictly on the worldvolume
terms in the action, has been extensively studied by recent authors
\cite{bi}.  In this section we consider the  complementary description of
extended objects in supergravity theories, which comes from focussing
on the `bulk' action terms, which describe the fields
which propagate in the bulk away from the brane.  These bulk terms
are of course just the effective supergravity Lagrangian terms which are 
obtained
from the low energy limit of string theory and/or M-theory.
One may therefore approximate the gravitational fields of $p$-branes,
at least semi-classically, by looking for solutions of the supergravity
equations of motion with the relevant symmetries.  

Generically, these
solutions will have event and Cauchy horizons, and there will no longer
exist any `brane worldvolume'.  A natural question, then, is where the
degrees of freedom associated with the brane are located.
Before tackling that question we shall consider some examples where
the location of the brane is relatively unambigous.

One of the simplest such self-gravitating brane solutions
is the 6-brane of eleven-dimensional supergravity.
Geometrically this is a product
\begin{equation}
{\cal E} \equiv M_{TN_k} \times {\Bbb E}^{6,1},
\end{equation}
where $M_{TN_k}$ is the multi-Taub-NUT metric
with $k$ centres
\begin{equation}
ds^2= V^{-1} (d\tau + \omega _i dx^i)^2 + V dx^i dx^i,
\end{equation}
with $V=1 + \sum { 1\over |{\bf x}- {\bf x}_i | }$. 
The group $G= U(1)$.  The 6-branes are located
at the ${\bf x}= {\bf x}_i$. These are fixed point sets
of the the Killing field $\partial / \partial \tau $
and hence, by a standard result, totally geodesic submanifolds.
Not only does the Killing field vanish on the branes but
so does the two-form (\ref{Killer}). 

Consider now two orthogonally intersecting sets of 6-branes.
 Geometrically we have
the product
\begin{equation}
{\cal E}\equiv M^{x} _{TN_k} \times M^{x^\prime} _{TN_{k^\prime} }\times 
{\Bbb 
E}^{2,1}.
\end{equation}
There are now two Kaluza-Klein $U(1)$ Killing fields, i.e. $G=
U(1) \times U(1) ^\prime$.
One Killing field vanishes at
${\bf x}= {\bf x}_i$ and the other at ${\bf x}^\prime = {\bf x}^\prime_i$.
However, apart from at the intersection, one $U(1)$ Killing vector potential
and the associated two-form (\ref{Killer})
are non-vanishing on the 6-brane of the other type.  

We are now in a position to apply Nielsen's argument
as formulated in the previous section. Clearly the branes at ${\bf x}^\prime = 
{\bf x}^\prime_i$ are 
superconducting with respect to $\partial / \partial \tau $
and the branes at ${\bf x} = {\bf x}_i$  are superconducting with respect to
$\partial / \partial \tau ^\prime $. In other words in this situation
each type of brane is superconducting with respect to the {\it other}
$U(1)$. Later we shall see a similar phenomenon arising in the case 
of extreme black holes in theories with two $U(1)$'s.  
 
The example we have just given may be readily extended to
the case of configurations
of branes intersecting at angles discussed in \cite{GGPT}.

So far we have not used the Einstein equations.
To do so, we suppose that the Killing vector field $K$ is everywhere
tangent to
some  submanifold $\cal B$ 
of $\cal E$.  We may regard $K$ as a Killing field of $\cal B$.
Of course $\cal B$ could be all of the spacetime manifold.

We  now apply the Ricci identity
to the Killing vector field $K$ to 
give
\begin{equation}
\nabla _i F^{ij} = - R_{\cal B} ^{ij}  K_j,
\end{equation}
where $R_{\cal B} ^{ij}$ \footnote{Our conventions are that the
signature is $(- + + ... +)$, and that the sign of the curvature is given by
$({\nabla}_{i}{\nabla}_{j} - {\nabla}_{j}{\nabla}_{i})K^{m} = 
{R_{nij}}^{m}K^{n}$.} is the Ricci tensor of $\cal B$.
Thus on $\cal B$ we have the London-like relation:
 \begin{equation}
J^i = 2R_{\cal B}^{ij} K_j.
\end{equation}
Moreover
\begin{equation}
\nabla ^2 K^i = - R_{\cal B}^{ij} K_j 
\end{equation}
As an example: suppose that the 
spacelike  submanifold ${\cal B}$ is spacelike,
compact and 
has negative Ricci curvature, then 
a simple integration by parts argument shows that
$K$ must  vanish everywhere on $\cal B$.
If $\cal B$ is Ricci-flat then $K$ need not vanish but if it does not
then it must be covariantly constant. This means that  locally at least
$\cal B$ is the
metric product of a circle with a submanifold of one dimension less
than that of $\cal B$. 

The result we have just sketched is responsible for the well
known fact that closed Einstein manifolds with negative cosmological 
constant do not admit any Killing fields. However 
we would like to view 
it in a a different way.

If $K$ vanishes on $\cal B$ then necessarily  the restriction
to $\cal B$ of $F=dK$ must also vanish. Thus the  submanifold $\cal B$
might be said to exhibit a kind of Meissner effect. 
Because the mathematical
theorem we are appealing to is a particular case of a Bochner 
vanishing theorem
it seems appropriate to refer to this effect as the Bochner-Meissner
effect. 

We now turn to spacetimes with event horizons.

Clearly the brane is located 
somewhere in the vicinity of the horizon. For a generic non-dilatonic 
$p$-brane,
the near horizon geometry is a standard compactification of
the form $(adS)_{p+2} ~{\times}~ S^{{d_T}-1}$, where $d_T$ is the 
dimension of the transverse space \cite{gazpaul,ght} (far from the brane the 
geometry is 
usually asymptotically flat, unless some global identification has been
performed).

Now the metric on $(adS)_{p+2}$  may be 
written in so-called {\it horospherical} coordinates $(t, {\bf x}_{p}, z)$:
\begin{equation}\label{adshor}
ds^2 = \frac{1}{z^2}[-dt^2 + d{\bf x}_{p}d{\bf x}_{p} + dz^2]
\end{equation}
These coordinates then provide a foliation of $(adS)_{p+2}$
by flat timelike hypersurfaces $z = $constant, which are called the
`horospheres'.  If one embeds $(adS)_{p+2}$ as a 
quadric in ${\Bbb E}^{p+1,2}$ then the horospheres are the 
intersection of the quadric with a family of null hyperplanes.

(The notation here reflects the fact that in the case
of  hyperbolic
space $H^{p+2}$, which  is the Euclidean section of 
adS, the analytic continuation of
the constant $z$ slices of (\ref{adshor}) are literally flat spheres, termed 
horospheres in the mathematics literature years ago. 
If one regards $H^{p+2}$ as the mass-shell in
$p+3$-dimensional Minkowski spacetime ${\Bbb E}^{p+2,1}$
then  horospheres are also the intersections of the quadric with
a family of null hyperplanes).

Now each horosphere may be thought of as a static
test test $p$-brane
which solves the Dirac-Born-Infeld
equations of motion of a $p$-brane coupled
to the $p+1$ potential $A^{p+1}$ whose $p+2$
 field strength $F^{p+2} =d A^{p+1}$ is proportional
to the volume form of $(adS)_{p+2}$ \cite{CKKTP}. In this way we
obtain a a particularly vivid picture of how the
heavy supergravity brane is composed of many stacked light branes. 

The limiting brane as $z\rightarrow 0$ corresponds to the causal boundary of 
$(adS)_{p+2}$. 
This conformal boundary has the topology of $S^{1} ~{\times}~ S^{p}$,
where the $S^1$ is the timelike factor and the $S^p$ is spacelike.
In fact the  boundary coincides (possibly up to a discrete
identification) with the conformal
compactification of  $p+1$ dimensional
 Minkowski space ${\Bbb E}^{p,1}$ and 
the isometry group $SO(p+1, 2)$ of $(adS)_{p+2}$ acts by conformal
transformations on the boundary.  Thus, one is led to study the 
singleton and doubleton representations \footnote{Singleton representations
of the adS group require a {\it single} set of oscillators transforming
under the fundamental representation of the maximal compact subgroup of
the covering group of the adS group; doubletons require two such sets of
oscillators.} of the group $SO(p+1, 2)$,
in the hope of understanding the conformally invariant quantum field
theory on the boundary.  In fact, this boundary QFT has precisely the
same degrees of freedom as the worldvolume fields of the corresponding
$p$-brane.  A natural proposal is then that the lowest scalar
component of the boundary field theory represents the transverse
fluctuations of the $p$-brane.  Indeed, most recently it has
been conjectured \cite{juan} that information about the
dynamics of superconformal field theories (in the large $N$ limit)
may be obtained by studying the region near the horizon of certain
D(p)-branes.  Thus, the conjecture implies a correspondence between
gauge theories in the large $N$ limit and compactifications of supergravity
theories.  The correspondence is often called `holographic' \cite{ed} because 
the
superconformal field theory (SCFT) lives on the causal boundary of adS.  

It is now natural to propose that a gravitating
$p$-brane is `superconducting' if the field theory on the boundary 
of the adS factor of the near-horizon geometry exhibits
behaviour characteristic of a superconducting phase.  Typically,
given any specimen
in a superconducting phase we expect to find zero modes, i.e., 
minimally coupled eigenmodes of some wave operator which correspond to 
the unimpeded movement of charge in the medium.  Thus, we are led
to look for zero modes which `skim along' the horospheres in the adS
factor.

{}From what we have said above, it is natural to look for such zero modes
in the singleton (or doubleton) supermultiplets.  After all, the 
singleton (or doubleton) field theories generically contain a number
of massless scalar and spinor fields, which are trapped on the boundary
of adS (the `core' of the brane).  (For an explicit discussion of the
matter content of the superconformal multiplet of the M5-brane see e.g.
\cite{ckvp}).  The precise form of these multiplets is not important.
What is important is that these massless modes skimming along the 
horosphere at infinity will naturally couple to any Kaluza-Klein
currents on the brane.  Put another way, if we wrap the brane on a 
circle (taking care to avoid any fixed-point singularities \cite{wrap}),
then the massless fermions on the dimensionally reduced brane will 
naturally couple to the Kaluza-Klein charge - these modes will 
induce a superconducting current on the reduced brane.

We are thus led to a pleasing microscopic description of the 
superconducting properties of self-gravitating branes.  Since the
supercurrent seems to live right at the horizon of the brane, we would
expect the horizon to display the Meissner effect.  In the next few
sections we will present a number of examples which confirm this 
effect for the horizons of extreme black holes.  It would be interesting to
perform similar tests for higher dimensional extremal self-gravitating branes.

Of course, all of this structure will break down for {\it non}-extreme
black branes.  As you approach the outer horizon, there is no splitting
of the spacetime geometry into an adS factor and a compact factor.
Furthermore, it is not possible to think of a non-extreme black brane
as a stack of light branes, all hovering just outside of the horizon.
We would not expect the outer horizon of a non-extreme brane to support
a superconducting current, and therefore we would not expect such an object
to display superconducting properties.  These expectations are borne out
when we consider non-extreme black holes.  It is always possible to 
penetrate non-extreme black hole horizons with magnetic flux;
superconductivity, it seems, is generically broken whenever we
break extremality.

\section{Meissner effect for superconducting strings}\label{strmeiss}

In the previous sections we have seen that the worldvolume of $p$-branes 
behaves like a superconducting medium with respect to gauge fields of 
Kaluza-Klein origin. In particular, a form of the London equation appears which 
implies the possibility of stationary currents in the absence of an external 
electric field. Another consequence of these macroscopic equations is that 
magnetic fields vanish inside the worldvolume, i.e., the 
Meissner expulsion of magnetic fields. As a matter of fact, the magnetic fields 
have vanishing normal component to the worldvolume. Of course, 
in order for the magnetic field to be interpreted as a vector field, we must 
restrict ourselves to four spacetime dimensions.

When the effects of self-gravitation are included, it becomes less clear where 
the brane is localized. Thus, it is not so evident where the Meissner surface, 
where magnetic field expulsion takes place, should be located. The arguments in 
the previous section suggest that, at least for non-dilatonic branes this 
should be in the near-horizon $adS$ throat. Dilatonic branes are singular at 
the horizon, and the $adS$/SCFT correspondence becomes less clear, but the 
singular horizon (or the close vicinity of it) would be the natural place for 
the brane. In this and the following sections we will argue that the Meissner 
surface is always precisely at the horizon.

The reader may feel that there is an apparent conflation of objects of 
different dimensionalities here. Consider a string, which we will wrap 
on a circle in the Kaluza-Klein fashion. The worldvolume viewpoint 
of the previous sections would lead to the conclusion that the string carries a 
superconducting current along itself. In the reduced spacetime the string 
worldvolume will look 
like a point, and it does not make much sense to speak about the field being 
expelled from a point. However, when we include gravity in the picture, the 
string 
will develop a horizon, which (in $D=4$) will be seen as a 2-sphere (the fact 
that this might be singular will be dealt with later). Our claim is that 
magnetic Kaluza-Klein fields are expelled from the horizon.

Hence, our starting point is a string in $D=5$ which is wrapped to yield a 
black hole. The 
metric, in Einstein frame in $D=5$ is
\begin{equation}\label{gravstring}
ds^2= H^{-1/3}(-f dt^2 +dz^2)+H^{2/3}\left({dr^2\over f} +r^2d\theta^2 
+r^2\sin^2\theta d\varphi^2\right),
\end{equation}
where
\begin{equation}
H=1+{q\over r},\qquad f=1-{r_0\over r}.
\end{equation}
For $r_0\neq 0$ there is an event horizon at $r=r_0$. When $r_0=0$ the string 
is extremal. 

If we compactify this geometry along the string direction $z$ we obtain a 
dilatonic black hole solution in $D=4$. In the previous sections we have seen 
that the string is superconducting with respect to the Kaluza-Klein gauge field 
${\cal F}$ generated along this isometry \footnote{In order to avoid confusion 
with other gauge fields that may appear, throughout 
this and the following sections we will consistently use caligraphic letters 
for the field ${\cal 
F}$ that experiences the Meissner expulsion and its potential ${\cal A}$.}. 
Our aim is to show that the horizon behaves as a Meissner surface for this 
field in the 
extremal limit.

There is an obvious point of concern when dealing with the extremal limit of 
the solution (\ref{gravstring}): the proper size of the 
horizon is zero as measured in the Einstein frame. However, in four 
dimensions the gauge field equation is conformally invariant. This means that 
the field does not distinguish whether we are working in the Einstein, string, 
or any other conformal frame related to the one above by an overall rescaling 
of the metric by a factor of the dilaton. In particular, there exists a a 
frame, namely $H^{4/3} ds^2$, in which the metric does not 
become singular at the horizon. In this frame it makes perfect sense to 
consider whether the field penetrates or not the horizon.

\begin{figure}
\begin{center}\leavevmode  %
\epsfxsize=14cm 
\epsfbox{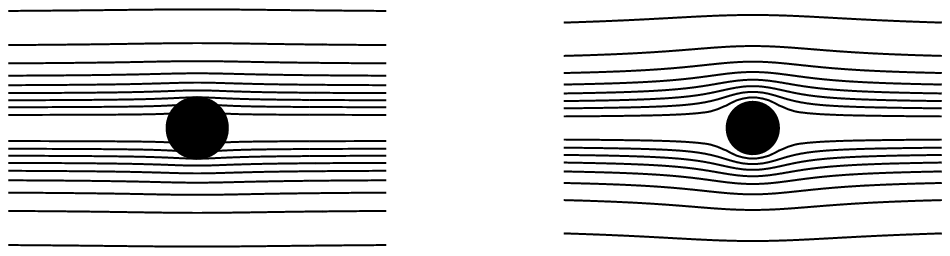}
\end{center}
\caption{Field lines of the Kaluza-Klein magnetic field ${\cal F}$ for the 
exact solution
(\ref{melvstr}), for the black holes that result from compactification of 
non-extremal and extremal strings. The radius in 
(\ref{melvstr}) has been changed to `Schwarzschild 
radius' $r\rightarrow r-q$.
}
\label{fig:MeissMelv}
\end{figure}

There is a well-known procedure to generate, upon reduction, an exact solution 
with an axisymmetric magnetic Kaluza-Klein field (see, e.g., \cite{dilatonc} or 
\cite{dggh}). Instead of identifying points along the orbits of 
$\partial/\partial z$, we twist the compactification direction to be along 
orbits of
\begin{equation}\label{twistkk}
{\bf q} ={\partial\over \partial z} +B{\partial\over \partial \varphi}.
\end{equation}
This is most easily done by changing to the adapted coordinate $\varphi 
\rightarrow
\varphi-Bz$, such that ${\bf q}\varphi=0$. Here $B$ will be the 
asymptotic value of the magnetic field along the axis of the tube. The 
Kaluza-Klein gauge potential ${\cal A}_\mu$ reads, in terms of the original 
metric,
\begin{equation}\label{harkk}
{\cal A}={q_{\varphi}\over |{\bf q}|^2}d\varphi=B{g_{\varphi\varphi} \over 
g_{zz}+B^2 g_{\varphi\varphi}}d\varphi.
\end{equation}
This is clearly a conformally invariant expression. For the case under 
consideration,
\begin{equation}\label{melvstr}
{\cal A}=B{H r^2\sin^2\theta\over 1+B^2 H r^2\sin^2\theta}d\varphi.
\end{equation}
We want to find the magnetic flux across a portion $\Sigma$ of the black hole 
horizon. This is given by the line integral $\int_{\partial \Sigma} {\cal A}$ 
on the 
horizon. If the horizon is at $r=r_0\neq 0$ then we find a non-vanishing flux 
across any portion of it. But in the extremal limit 
the horizon is at $r=0$, where ${\cal A}$ vanishes. So no magnetic flux 
penetrates the extremal horizon. The field is expelled from it: this is the 
Meissner effect. In Fig.~\ref{fig:MeissMelv} we have plotted the lines of force 
of the magnetic field for non-extreme and extreme configurations. 

We would like to emphasize the fact that this analysis has been carried out at 
a level where the supergravitry equations have been treated in an exact form. 
In particular, the field (\ref{melvstr}) is an exact field configuration in 
$D=4$ (together with the corresponding metric and string winding field).

\section{Meissner effect in extremal black holes}\label{extrmeiss}

It is remarkable that this Meissner effect is not unique of
extremal geometries derived from $p$-branes. In fact, as we argue below, it 
appears to be a rather generic feature of extremal black holes. Typically, 
the lines of force of a magnetic field penetrate the horizon of a non-extremal 
black hole. However, we will see that
the lines of force fail to penetrate extremal horizons.
Instead, they tightly wrap the black hole. 
{\it The horizon of an extremal black hole behaves 
like the surface of a perfectly diamagnetic object.} 

To be more 
precise, in a superconducting material the magnetic field penetrates to some
small distance from the surface: this is the penetration depth. For extremal
black holes the penetration depth appears to be zero. Also, the perfectly
diamagnetic
state of the black hole breaks down at any finite temperature, i.e., for
any deviation from extremality. 

To our knowledge, this phenomenon 
was first pointed out in the literature by Bi\v{c}{\'a}k and Dvo\v{r}{\'a}k 
in \cite{bicak}, in the context of Einstein-Maxwell theory. We believe this 
to be a generic phenomenon for black holes in theories with more complicated 
field content, although a precise specification of the dynamical situations 
where this effect is present seems to be out of reach. The results below 
constitute very strong evidence that it is true whenever the gauge field 
couples minimally to the geometry, or possibly includes dilatonic couplings.

\subsection{Field expulsion from extremal rotating black hole}\label{walds}

A first example (also noticed in \cite{bicak}) of this Meissner effect 
follows from Wald's analysis 
\cite{wald} of
a test magnetic field in the background of the neutral Kerr black hole.
In \cite{wald} a solution for a field aligned with the axis of the black hole 
is constructed, by using the isometries of the Kerr background. Let us denote 
the axial and temporal Killing vectors of the Kerr solution by $\psi \equiv 
\partial /\partial\varphi$ and  $\eta \equiv \partial /\partial t$. Then a test 
gauge field can be constructed as
\begin{equation}
{\cal A}_\mu= B\left( \psi_\mu + {2 J\over M}\eta_\mu \right) -{Q\over 2 
M}\eta_\mu.
\end{equation}
$B$ is the magnetic field along the axis, and $Q$ is the charge that the black 
hole acquires, which we want to be zero. The field can be conveniently written 
in terms of the vector $\chi =\Omega_H \psi +\eta$, which is tangent to the 
null geodesic generators of the horizon. Here $\Omega_H$ is the angular 
velocity of the horizon. We find (with $Q=0$),
\begin{equation}
{\cal A}_\mu= {B\over \Omega_H}\left[ \chi_\mu - \left(1-{2 \Omega_H J\over 
M} \right)\eta_\mu\right].
\end{equation}
In the extremal limit $2\Omega_H J=M$, and therefore ${\cal A}_\mu \propto 
\chi_\mu$, which vanishes precisely at the horizon. As in the previous 
section, the flux along any portion $\Sigma$ of the horizon, $\int_{\partial 
\Sigma} 
{\cal A}$, vanishes. Again, the extremal horizon behaves like a perfect
diamagnet.

This solution involved the magnetic field as a test field only. But it is 
possible to find an exact generalization of it within Kaluza-Klein theory.
Start with the product of the (neutral) 
$D=4$ Kerr solution with
a five dimensional direction $x^5$.
We can now apply the `twisted reduction' procedure described in Section 
\ref{strmeiss} to put the $D=4$ neutral Kerr black hole 
in the background of an axisymmetric Kaluza-Klein magnetic field in an exact 
way. In order to avoid the presence of electric charge in the black hole, the
compactification direction must also involve a twist in the time coordinate. 
Specifically, we identify points along orbits of the vector
\begin{equation}
{\bf q} = {\partial\over \partial x^5}+B\left( \psi + {2 J
\over M}\eta \right).
\end{equation}
The {\it exact} Kaluza-Klein gauge field that follows is
\begin{equation}
{\cal A}_\mu =B{\psi_\mu + {2 J \over M}\eta_\mu \over |{\bf q}|^2},
\end{equation}
which reduces to Wald's field in the linear approximation, and in the same way 
can be seen to exhibit the Meissner effect in the extremal limit.
The reader may have noticed that Wald's solution does not contain any
dilaton field, whereas the Kaluza-Klein solution does. But to 
linearized order in the
test gauge field there
is no contribution from a test dilaton (see, e.g., eqn.~(\ref{testsc}) below). 
Therefore
Wald's solution is the linear approximation to the axial field configuration
for {\it all} Einstein-Maxwell-dilaton theories \footnote{Actually, the 
Kaluza-Klein perspective provides a simple way to rederive, by linearization
in the gauge field, the general technique used in \cite{wald} to construct
solutions for test Maxwell fields in backgrounds with isometries.}.

Finally, in the solutions we have been considering the magnetic field is 
aligned with the rotation axis of the black hole. According to \cite{bicak}, 
the Meissner expulsion can also be seen for fields where no alignment is 
assumed.

\subsection{Field expulsion from spherically symmetric extremal 
throats}\label{nrthroats}

Now we would like to consider other classes of extremal
black holes, and the most obvious candidates are charged (Reissner-Nordstrom)
black holes. However, several subtleties arise that need to be dealt with
care. Consider, as the simplest example that comes to one's mind, an 
electrically charged Reissner-Nordstrom black hole in the background
of a magnetic field. This configuration was analyzed, in an exact way, in
\cite{ernst}. Naively,
according to our conjecture the magnetic field should be expelled 
from the horizon
in the extremal limit in this configuration. However, this does not happen. The 
puzzle is solved 
\cite{bicak} when
one notices that the solution in \cite{ernst} is actually {\it rotating.}
A rotating electric charge generates a magnetic dipole moment. The black hole
is therefore the source of a magnetic dipolar field. This is actually 
the field across the extremal horizon of
the solution in \cite{ernst}\footnote{It is even clearer that, for similar
reasons, we should not expect the extremal Kerr-Newman black hole, which
has a magnetic dipole by itself, to 
expel the magnetic field \cite{ew}.}. The authors of \cite{bicak} then went 
on to construct a linearized solution where the rotation of the charged black 
hole in the external field could be set to zero, and found it exhibited the 
Meissner expulsion of the field in the extremal limit.

In this example, the complication arises due to gravitationally induced 
non-linear interactions
between the electric field of the black hole and the external magnetic field.
However, notice that our main reason to have a charge on the black hole is
to provide a means to reach the extremal limit. In other words, we are not
particularly interested in the dynamical aspects associated to the charge of 
the black hole. Rather, we want to isolate the behavior of the magnetic field 
in the
{\it gravitational} field created by the black hole. 
As a way to disentangle the effect of the charge of the black hole from that
of the magnetic field, 
we can think of the charge of the black hole as being coupled to a 
gauge field that is different from the
external magnetic gauge field.
In other words, we work with a
$U(1)\times U(1)$ gauge theory, with two Maxwell fields.
The black hole will be charged with respect to one of the $U(1)$ fields, 
while
the other gauge field will be the magnetic field that experiences 
the Meissner effect.
This introduction of a second gauge field may seem unrealistic, but we should
view it as simply a device that provides us with a way to achieve extremality
for the black hole. In particular, it will be clear in our analysis below
that the dynamics of the
gauge field associated to the charge of the black hole play no essential 
role.
Besides, theories with more than one gauge field arise quite naturally in
string theory and related contexts.

We will start our analysis by treating the magnetic field as a test field
in the background of the black hole geometry. Therefore, we want to solve the 
equation
\begin{equation}\label{testf}
\partial_\mu (\sqrt{-g} {\cal F}^{\mu\nu})=0,
\end{equation}
in some fixed background geometry $g_{\mu\nu}$.

For starters, take the Reissner-Nordstrom metric,
\begin{eqnarray}
ds^2 &=& -V dt^2 + V^{-1} dr^2 +r^2 (d\theta^2 +\sin^2\theta d\varphi^2),\\
V &=& 1-{2 M\over r} + {Q^2\over r^2}.\nonumber
\end{eqnarray} 
The outer (event) horizon is at $r=r_h=M+\sqrt{M^2-Q^2}$, and extremality is
achieved by setting $Q=M$.

For the test field ${\cal F}=d{\cal A}$ we will assume the ansatz
\begin{equation}\label{Aansatz}
{\cal A} = f(r) \sin^2\theta\; d\varphi,
\end{equation}
in terms of which the magnetic flux crossing any surface $\Sigma$
is given by $\int_{\partial \Sigma} {\cal A}$.

With the ansatz (\ref{Aansatz}), the field equation (\ref{testf}) becomes
\begin{equation}
{d\over d r}\left( V{d f\over d r}\right) =
{2 f\over r^2}.
\end{equation}
This is easily solved as
\begin{equation}
f(r) = r^2 -Q^2,
\end{equation}
up to a multiplicative constant, related to the value of the magnetic field 
at 
infinity, which we have arbitrarily fixed. According to (\ref{Aansatz}), the
magnetic flux crossing the horizon is proportional to $f(r_h)$. This is 
nonzero for black holes with $M>Q$, but {\it it vanishes precisely in the
extremal limit
$r_h = Q$.} 

Now, we want to consider non-rotating extremal black holes in more 
generality.
In order to simplify the analysis, we will focus only on the region 
near the horizon of the black hole, since it 
is there where the Meissner effect is exhibited.
As the most generic characterization of this region for spherically symmetric 
extremal black holes, we will take the following:

\begin{itemize}

\item For some choice of conformal frame, the region near the extremal 
horizon
becomes asymptotically an infinite throat of constant radius. 
This is, if we choose
the horizon to be at $r=0$, then
\begin{equation}\label{exthroat}
ds^2 \simeq -\left( {r\over \ell}\right)^{4\alpha} dt^2+
\ell^2\left[ {dr^2\over r^2} + d\theta^2 + \sin^2\theta d\varphi^2
\right].
\end{equation}

\end{itemize}

The freedom in choosing coordinates has been used to simplify the possible
forms of the
metric and bring the horizon to $r=0$. 
The parameter $\ell$ fixes the scale of the geometry (and is tipically 
related
to the charge and mass of the black hole). The exponent $\alpha$ is an
arbitrary real number. Within this class we find, for example, the 
extremal dilatonic 
black holes of \cite{gm}, or the stringy black holes in \cite{cvetyoum2}. 

As in Section \ref{strmeiss}, the reference to the conformal frame is motivated 
by the fact that, in the
presence of scalar (dilaton) fields, when we write the metric in canonical 
Einstein frame, the throat at $r=0$ typically pinches down to zero size
in a singular way. But then we can use the dilaton to perform a conformal 
rescaling of the metric to yield the regular throat (\ref{exthroat}).
Since the Maxwell field equation (\ref{testf}) is, in four dimensions, 
invariant under such conformal rescalings, we are allowed to choose to
work in the conformal gauge fixed by (\ref{exthroat}). In fact, we may want
to consider an equation slightly more general than (\ref{testf}), 
\begin{equation}\label{testfd}
\partial_\mu (\sqrt{-g} e^{-a\phi} {\cal F}^{\mu\nu})=0,
\end{equation}
where we allow
for a coupling of the test field to a dilatonic field $\phi$ with 
non-constant 
background value near the horizon, $r\simeq 0$,
\begin{equation}
e^{-a\phi} \simeq B\left({\ell\over r}\right)^{2\beta}.
\end{equation}

As a further minor generalization, we could consider the test gauge 
field to be coupled to a {\it test} scalar $\sigma$, with the 
standard action (we suppress inessential factors),
\begin{equation}\label{testsc}
I \sim \int (\partial \sigma)^2 + e^{-\sigma} {\cal F}^2.
\end{equation}
However, the field
equation for $\sigma$
implies that if ${\cal F}$ is linear in the (small) applied magnetic field, 
then
$\sigma$ only enters at quadratic order and is therefore negligible
in the approximation we are working. Hence we need not consider 
explicitly such scalars.

In order to solve (\ref{testfd}), we consider again the ansatz 
(\ref{Aansatz}) for the magnetic field, and we
find the equation
\begin{equation}
{d\over dr}\left(r^{2(\alpha-\beta) +1} f'\right) = 
2 r^{2(\alpha-\beta)-1} f.
\end{equation}
This is a homogeneous equation, which we can solve by choosing (up to a
multiplicative constant), 
\begin{equation}\label{ggam}
f(r) = r^\gamma,
\end{equation}
with
\begin{equation}\label{gammasol}
\gamma = \sqrt{(\beta -\alpha)^2 + 2} +\beta-\alpha >0.
\end{equation}
What is important here is that $\gamma$ is never zero. 
Since the flux crossing the horizon
is proportional to $f(r=0)$, in order to have a finite, non-vanishing flux
we should have $\gamma =0$. Instead, we find that the flux always vanishes
at the horizon $r=0$ \footnote{The solutions with $\gamma <0$ have been 
discarded as pathological.}. 
The Meissner effect, therefore, is a
common characteristic of extremal throats. For completeness, we show in 
Appendix \ref{nonmeiss} that the Meissner effect does never take place on 
non-extremal horizons.

Finally, notice that in order to solve the equations and exhibit the Meissner
effect we have only needed the metric of the black hole solution. That is,
the fact that we may need the black hole 
to be charged for it to be
extremal, plays no essential role.
Besides this, we have assumed that the interactions of
the gauge field ${\cal F}$ are essentially given by (\ref{testfd}). More
complicated situations could be envisaged, but from the evidence we have 
presented here we believe that the phenomenon is generic. 
If other couplings of the field ${\cal F}$ were considered, care should be 
exercised to ensure
that the additional interactions do not indirectly generate source terms for
the field ${\cal F}$, which would produce an outgoing
flux of the field across the horizon. These cases, of course, cannot be used
to disprove our conjecture, which clearly requires absence of magnetic 
sources inside the black hole. A subtle example of how flux 
can penetrate a horizon of 
the type (\ref{exthroat}), if the theory contains Chern-Simons couplings 
involving
the field ${\cal F}$, is discussed in Appendix \ref{nonsuc}.

\subsection{Some further exact solutions}
\label{exactsols}

In the previous subsection we have found evidence that magnetic fields are 
expelled from the horizon of spherically symmetric extremal black holes. 
However, the magnetic field
has been treated as a test field, and its effect on the geometry of the black
hole has been neglected. One could worry that, if the backreaction 
effect of the magnetic field on the geometry were accounted for, the 
behavior of the horizon might change and the magnetic field would perhaps
penetrate into the black hole, thereby evading the Meissner effect. This, 
however, is rather unlikely: the fact that the magnetic field vanishes near the 
horizon leads us to expect a negligible backreaction in that region.
This expectation is confirmed in all cases where exact solutions have been 
constructed. 

We have already presented two exact solutions, in Sections \ref{strmeiss} and 
\ref{walds}, using the Kaluza-Klein ansatz, where we have introduced an 
axisymmetric magnetic field which exhibits Meissner expulsion. Similar exact 
fields can be introduced, for different values of the dilaton coupling, by 
applying `Harrison-like' \cite{harrison} solution-generating transformations 
\cite{dilatonc,ross,yo} (dilatonic Melvin flux tubes were discussed in 
\cite{gm}). In particular, the behavior of black holes in magnetic fields, for 
essentially any value $a>0$ of the dilaton coupling, can be readily analyzed 
using the solutions in \cite{yo}.
We will not give any details, but in all such cases the Meissner effect can be 
seen to be present as well. Here we will display another sort of magnetic 
fields that can, in a sense, be considered as a curved 
space generalizations
of the uniform magnetic field in flat space. These are the covariantly 
constant fields, exemplified by the Bertotti-Robinson solution of
Einstein-Maxwell theory. There do exist generalizations of such solutions 
for the $U(1)^2$ theory of \cite{klop}, or the $U(1)^n$ theories in \cite{yo}.

One should be careful, however, in constructing the solutions. The field
in the Bertotti-Robinson solution is spherically symmetric, and `emanates' 
from an origin, which nevertheless
is non-singular since the geometry develops an infinite throat. In the
analogous dilatonic solutions, the field similarly emanates from an origin, 
which now is singular in Einstein frame. In any case, our point here is that,
if we want the extremal black hole to expel the field, then it is clear that
the `source' should not be {\it inside} the black hole. In other words,
the Bertotti-Robinson-like field and the black hole must {\it not} be
concentric.

With this proviso, the theory we will consider will be \cite{klop}
\begin{equation}\label{u12}
I=\int d^4 x \sqrt{-g} \left[
R-{1\over 2}(\partial \phi)^2 -{e^{-\phi}\over 2} {\cal F}^2 
-{e^\phi\over 2}  G^2
\right],
\end{equation}
and the solution we are interested in is, in Einstein 
conformal gauge,
\begin{eqnarray}\label{bhbr}
ds^2 &=& -{1\over \Delta_{\cal F}\Delta_G}dt^2 + 
\Delta_{\cal F}\Delta_G (dr^2 +
r^2d\theta^2 +r^2\sin^2\theta d\varphi^2),\nonumber
\\
{\cal F}&=&d{\cal A},\quad {\cal A} =b {r\cos\theta-\ell\over 
r_2}d\varphi,\qquad G =
q \sin\theta d\theta \wedge d\varphi,
\\
e^{\phi} &=& {\Delta_G\over\Delta_{\cal F}},\quad \Delta_G = 1 +{q\over 
r},\qquad \Delta_{\cal F} = {b\over r_2},
\nonumber\\
r_2&\equiv &\sqrt{r^2 +l^2 -2lr\cos\theta}.\nonumber
\end{eqnarray}
In this form of the solution, both fields are of magnetic type. The black 
hole
is extremal from the outset, with horizon at $r=0$ and charge $q$. 
The `origin' of 
the magnetic ${\cal F}$
field is at a coordinate distance $l$ along the 
axis $\theta=0$, i.e., at $r_2=0$. 
Setting $q=0$ yields a geometry that is conformally 
equivalent to the
product of the linear dilaton vacuum of $D=2$ string theory 
with a sphere $S^2$, and a covariantly constant field ${\cal F}$. 
The degenerate horizon at $r0=0$ is singular. The proper 
size of the extremal black hole is zero if measured in the Einstein metric.
However, as discussed in previous sections, for the purpose of
studying the gauge fields
we could just as well work in a conformally related metric where 
the extremal horizon is non-singular. The `preferred' frame is $e^{\phi} 
ds^2$, in which the extremal black hole area is equal to $4\pi q^2$.

\begin{figure}
\begin{center}\leavevmode  %
\epsfxsize=6cm\epsfbox{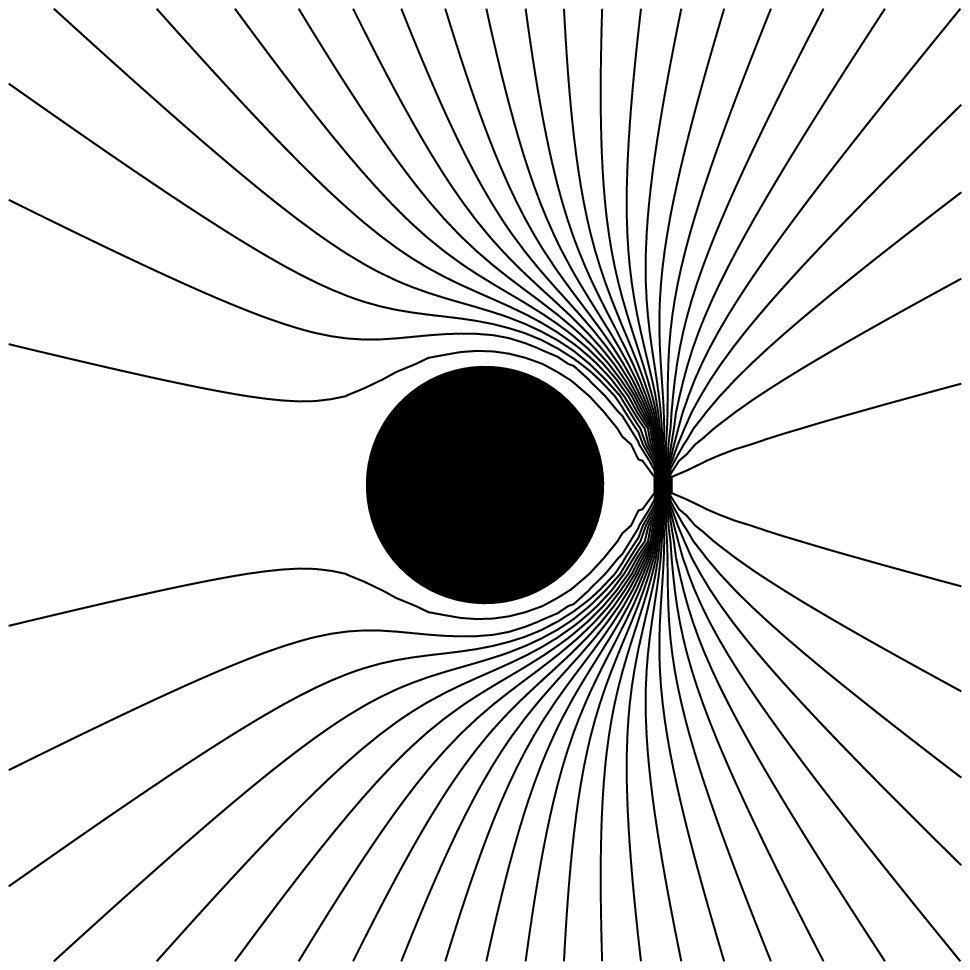}
\end{center}
\caption{Field lines of the magnetic field ${\cal F}$ for the exact 
configuration
(\ref{bhbr}). The radius in (\ref{bhbr}) has been changed to `Schwarzschild 
radius' $r\rightarrow r-q$. The `origin' of the covariantly constant field 
has 
been put at $l=q/2$.
}
\label{fig:MeissBR}
\end{figure}

Once again, the exact value of the flux across constant $r$ surfaces, given by 
\begin{equation}\label{bhbrf}
{\cal F}_{\theta\varphi} = 
b\sin\theta{r^2(r-l\cos\theta)\over r_2^3}
\end{equation}
vanishes at the horizon of the black hole, $r=0$, as we claimed. The lines of 
force 
for the field ${\cal F}$ are plotted in Fig.~\ref{fig:MeissBR}.

With little extra effort we can consider a slightly different situation,
where we have two extremal black holes, each with charge coupled 
to different gauge
fields. As before, if we do not want to find a trivial penetration
of flux, we have to consider a two-center solution.

We can analyze in this way whether the field created by the black hole
with charge $q_2$ in ${\cal F}$ penetrates the 
horizon of the black hole with charge $q$.
The solution is just like (\ref{bhbr}) above, but now with
\begin{equation}
\Delta_{\cal F} = 1 +{ q_2\over r_2}.
\end{equation}
The horizon of this second black hole is at $r_2=0$. The field created by
it is exactly the same as in the previous example, (\ref{bhbrf}), only 
changing $b\rightarrow  q_2$. Thus we find
another exact solution exhibiting the Meissner effect at the extremal
horizon at $r=0$.
Evidently, by symmetry, the flux created by the black hole with charge
$q$ does not penetrate the horizon, at $r_2=0$, 
of the other extremal black hole.

In these examples the black hole 
under study
has been the ``$a=1$ dilatonic black hole.'' In terms of the test field
analysis performed in the previous subsection, the relevant 
parameters are $\alpha=0$, 
$\beta =1/2$, which yield $\gamma = 2$ for (\ref{ggam}). This is in precise 
agreement with the expansion for small $B$ (and $r$) of 
the exact result (\ref{bhbrf}). Different values of the
dilaton coupling (essentially, any value $a>0$) 
can be readily analyzed using the solutions in \cite{yo}, with no qualitative 
differences.

\section{Conclusions}

Superconductivity is a rich and multifaceted subject, with applications
in a variety of physical models, from condensed matter physics to QCD.
It is therefore natural to investigate how superconducting
phenomena may emerge from the rich structure described by M-theory; after all,
M-theory is our only real candidate for a unified description of all
physical phenomena.

In this paper, we have described the superconducting phases of the solitonic
objects of M-theory, the $p$-branes.  In order to perform such a description, 
we have concentrated on three of the most elementary
and well known aspects of superconducting
media:  The Meissner effect, the London theory and the existence of 
minimally coupled zero modes.

With respect to the Meissner effect, we have presented a number of 
exact solutions which demonstrate that Kaluza-Klein magnetic flux is
expelled from the horizon of a generic extreme black hole.  We have 
extended this analysis to the case of a black string in $D=5$, and again
found that Kaluza-Klein flux is expelled.  It would be interesting to perform
similar tests for the Meissner effect for higher-dimensional extreme branes.
It would also be interesting if we could understand {\it precisely}
when and how the Meissner effect is broken.

Strictly speaking, the Meissner effect follows from the fact that inside a 
superconductor the field has to be pure gauge. This, however, is not true for 
the field in the interior of the extremal black hole, as can be readily seen 
from the examples above. We are not claiming, therefore, that the black hole 
interior is in a superconducting state. Our statements refer to the horizon, or 
at most to the near-horizon region.

Of course, the Meissner effect is just one property exhibited by 
superconducting
media; ultimately, we want to construct a phenomenological model which
attempts to describe what is going on.  The theory of London goes beyond
the simple {\it observation} of the Meissner effect, and proposes a set
of field equations which imply various things about the microscopic
theory which underlies the entire phenomenon.  Thus, in order to have
a macroscopic phenomenological description of a superconducting $p$-brane,
we have followed Nielsen, Balachandran and others by proposing that a
$p$-brane is in a superconducting phase if and only if the relativistic
London equation holds on the worldvolume of the brane.  For a test brane,
this definition is not ambiguous since it is clear where the brane is located,
i.e., the brane is just some extended object moving in a background spacetime,
from which it has decoupled.  The motivation for our definition is then clear,
since the London equation will hold on the worldvolume of any extended object
which is carrying Kaluza-Klein currents.  For self-gravitating branes, we 
have proposed that a brane is in a superconducting phase if and only if
this `Nielsen' type condition holds on the boundary of the adS factor of the
near-horizon geometry of the brane.   

Given all of this structure, it is then natural to propose that the 
microscopic degrees of freedom which lead to $p$-brane superconductivity
are precisely the zero modes, associated with the singleton superconformal
multiplets, which propagate on the boundary of the adS factor of the 
near-horizon geometry.  These zero modes naturally couple to any Kaluza-Klein
currents, and so they literally represent the unimpeded flow of charge far
down the throat of a self-gravitating brane.

Of course, in this analysis we have neglected a number of other theories
and approaches to superconductivity.  It would be interesting
to investigate whether or not it is possible to define $p$-brane
superconductivity using the ideas of these other theories.
Research on these and related problems is currently underway.

\acknowledgements

AC is supported by Pembroke College, Cambridge.
RE has been partially supported by EPSRC through grant GR/L38158 
(UK), and by UPV through grant 063.310-EB225/95 (Spain).

\appendix

\section{Absence of Meissner effect in non-extremal horizons}\label{nonmeiss}

In order to complete our general analysis of {\it test} magnetic fields in the 
vicinity of spherically symmetric black holes, here we solve the equations in 
the presence of non-extremal horizons. In this case, close enough to the 
horizon the geometry is of the Rindler form
\begin{equation}
ds^2 = -\rho^2 d\tau^2 + d\rho^2 + R^2(d\theta^2 +\sin\theta^2 d\varphi^2).
\end{equation}
$R$ is a constant measuring the radius of the horizon, which is at $\rho=0$.
We now solve the equation (\ref{testf}) for a test Maxwell field in this 
background using the same ansatz (\ref{Aansatz}) \footnote{We could also have 
included scalar fields, as in (\ref{testfd}), but these typically take finite, 
non-zero values on non-extremal horizons and do not alter the results.}. The 
solution
\begin{equation}
{\cal A}= I_0\left( {\sqrt{2}\rho \over R} \right) \sin^2\theta d\varphi,
\end{equation}
is expressed in terms of the Bessel function of order zero, such that $I_0(0)= 
1$, i.e., there is a non-vanishing flux crossing any portion of the horizon. 
There is no Meissner expulsion from non-extremal horizons.

\section{A `counterexample' to the Meissner effect, 
and its resolution}\label{nonsuc}

Consider the five-dimensional action
\begin{equation}
I_5 = \int d^5 x \sqrt{-\hat g} \left\{\hat R - {1\over 2} (\hat\partial
\phi)^2 -{1\over 12} e^{-\sqrt{2/3}\phi}\hat H^2 -{1\over 4}
e^{+\sqrt{2/3}\phi}\hat F^2 \right\}.
\end{equation}
Five dimensional quantities will be hatted. $\hat H$ and $\hat F$ are
3-form and 2-form field strengths, obtainable from 2- and 1-form
potentials $\hat B,\hat A$, $\hat H = d\hat B$, $\hat F = d\hat A$.
Very similar (but not exactly the same) actions can be derived from
compactified string/M-theory. The fields $\hat H$ and $\hat F$ admit the
interpretation of fields with string and particle sources. Actually, the 
solution we discuss below can be seen as a bound state (at threshold) of a 
string and a particle. 
 
The equations of motion of this theory admit the
solution
\begin{eqnarray}
d\hat s^2 &=& -{dt^2\over \Delta^2} + \Delta^2(dr^2 +r^2 d\Omega_2^2) 
+ dx_5^2,\\
\Delta &=&  1+{q\over r},\nonumber
\end{eqnarray}
\begin{equation}
\hat B = \Delta^{-1}dt \wedge dx_5,\qquad \hat A = \Delta^{-1}dt.
\end{equation}
The scalar $\phi$ is zero (or constant) for this solution \footnote{It would 
be easy to construct a 
more general solution with different harmonic functions $\Delta_F,\Delta_H$
for the particle and string,
that would yield non-constant $\phi$, but we prefer to keep things simpler
at this level.}.
The metric is precisely equal to the product of the $D=4$ extremal, electric 
Reissner-Nordstrom black hole with the real line $-\infty <x_5 <\infty$.
Hence, Kaluza-Klein reduction along $x_5$ yields the extremal
RN black hole, with no electromagnetic Kaluza-Klein field.

We can now generate a background Melvin flux tube by performing a 
Kaluza-Klein 
reduction as described in Section \ref{strmeiss}:
change the polar variable to $\varphi \rightarrow\varphi - B x_5$, and reduce 
to 
$D=4$ by consistently identifying points along $x_5$. The Kaluza-Klein gauge 
potential
is 
\begin{equation}
{\cal A} = B{\Delta^2 r^2 \sin^2\theta\over 1+B^2\Delta^2 r^2 
\sin^2\theta}d\varphi.
\end{equation}
This does {\it not} vanish on the extremal horizon $r=0$.
The Meissner effect is not present for this
solution. Nevertheless, the geometry near the horizon is of the form 
required in (\ref{exthroat}).

The resolution of this puzzle comes from examining the actual couplings of
the Kaluza-Klein gauge field ${\cal A}$ in the effective $D=4$ theory.
For details of the reduction procedure, see, e.g., \cite{stelle}. The 
important point here is that the non-vanishing component of the field $\hat 
B$
along $x_5$,
$\hat B_{\mu 5}\equiv B_\mu$, yields a Chern-Simons-like
coupling in the $D=4$ action of the form
\begin{equation}
(d B\wedge {\cal A})^2
\end{equation}
(times factors involving the scalar $\phi$ and Kaluza-Klein scalar,  which 
are inessential for this discussion). The consequence is that the 
effective equation for ${\cal F}$
in $D=4$ differs now from (\ref{testfd}) by the presence of an extra  
source term. In this indirect way, the $\hat H$-charge of the black hole 
is responsible for the appearance of an induced magnetic dipole for the black
hole in the presence
of an external field ${\cal F}$. 
This is the source of the flux coming out of the horizon. This is, in a way, 
similar to absence of Meissner effect in the 
solutions considered in \cite{ernst}, in that subtle non-linear interactions
induce dipolar sources for the black hole.

This extra term is also present in the compactification of the string that we 
analyzed in Section \ref{strmeiss}. However, in that case its value in the 
extremal limit is zero, so it does not spoil the Meissner effect.

\begin{thebibliography}{99}

\bibitem{wald} R. M. Wald, {\it Black hole in a uniform magnetic field},
Phys. Rev. D{\bf 10} (1974) 1680.

\bibitem{klk} A.R. King, J.P. Lasota, and W. Kundt, {\it Black holes and 
magnetic fields,} Phys. Rev. D{\bf 12}, 3037 (1975).

\bibitem{ernst} F.J. Ernst, {\it Black holes in a magnetic universe}, Jour.
of Math. Physics, 17, (1976) 54.

\bibitem{ew} F.J. Ernst and W.J. Wild, {\it Kerr black holes in a magnetic
universe}, Jour. of Math. Physics, 17, (1976) 182.

\bibitem{bicak} J. Bi\v{c}{\'a}k and L. Dvo\v{r}{\'a}k, {\it Stationary 
electromagnetic fields
around black holes. III. General solutions and the fields of current loops 
near
the Reissner-Nordstrom black hole}, Phys. Rev. D{\bf 22} (1980) 2933.

\bibitem{cham} A. Chamblin, J.M.A. Ashbourn-Chamblin, R. Emparan,
and A. Sornborger, {\it Absence of Abelian-Higgs hair for extreme black
holes}, Phys. Rev. Lett. {\bf 80}, 4378 (1998) (gr-qc/9706032); {\it Can 
extreme black holes have (long) Abelian Higgs hair?,}  gr-qc/9706004.

\bibitem{nielsen} N.K. Nielsen, {\it Dimensional reduction and classical
strings}, Nucl. Phys. {\bf B167}, 149 (1980).

\bibitem{nielole} N.K. Nielsen and P. Olesen, {\it Dynamical properties
of superconducting cosmic strings}, Nucl. Phys. {\bf B291}, 829-846 (1987).

\bibitem{bal} A.P. Balachandran, B.S. Skagerstam, and A. Stern,
{\it Gauge theory of extended objects}, Phys. Rev. D{\bf 20},
439, (1979). 

\bibitem{witten} E. Witten, {\it Superconducting strings,} Nucl. Phys. {\bf 
B249}, 557 (1985). 

\bibitem{bi} G. W. Gibbons, {\it Born-Infeld particles and Dirichlet 
$p$-branes},
Nucl. Phys. {\bf B514}, 603, (1998) (hep-th/9709027);
C. G. Callan, Jr. and J. M. Maldacena, 
{\it Brane Dynamics From the Born-Infeld Action}, Nucl. Phys. {\bf B513},
198 (1998) (hep-th/9708147).

\bibitem{GGPT} J.P. Gauntlett, G.W. Gibbons, G. Papadopoulos and P.K. Townsend,
{\it Hyper-Kahler manifolds and multiply intersecting branes},
Nucl. Phys.  {\bf B 500}, 133, (1997) (hep-th/9702202). 

\bibitem{gazpaul} G.W. Gibbons and P.K. Townsend, {\it Vacuum interpolation
in supergravity via super $p$-branes}, Phys. Rev. Lett. {\bf 71}, No. 23,
3754 (1993).

\bibitem{ght} G. W. Gibbons, G. T. Horowitz, and P. K. Townsend,
{\it Higher dimensional resolution of dilatonic black hole 
singularities}, Class. Quant. Grav. {\bf 12} (1995) 297 
(hep-th/9410073).

\bibitem{CKKTP} P. Claus, R. Kallosh, J. Kumar, P.K. Townsend, and A. Van 
Proeyen, {\it Conformal Theory of M2, D3, M5 and 'D1+D5' branes,} 
hep-th/9801206

\bibitem{juan} J.M. Maldacena, {\it The Large N Limit of Superconformal Field
Theories and Supergravity}, hep-th/9711200.

\bibitem{ed} E. Witten, {\it Anti-de Sitter space and Holography}, 
hep-th/9802150.

\bibitem{ckvp} P. Claus, R. Kallosh and A. Van Proeyen,
{\it M5-brane and superconformal (0,2) tensor multiplet in 6 dimensions},
Nucl. Phys. {\bf B 518}: 117-150, (1998) (hep-th/9711161). 

\bibitem{wrap} G.W. Gibbons, {\it Wrapping branes in space and time},
hep-th/9803206.

\bibitem{dilatonc} F. Dowker, J.P. Gauntlett, D. Kastor and J. Traschen,
{\it Pair creation of dilaton black holes},
Phys. Rev. D{\bf 49} (1994) 2909 (hep-th/9309075).

\bibitem{dggh} H. F. Dowker, J. P. Gauntlett, G. W. Gibbons, and
G. T. Horowitz, {\it Nucleation of $p$-branes and fundamental strings},
Phys. Rev. D{\bf 53} (1996) 7115 (hep-th/9512154).

\bibitem{gm} G. W. Gibbons and K. Maeda, {\it Black holes and membranes
in higher-dimensional theories with dilaton fields,} Nucl. Phys.
{\bf B298} (1988) 741.

\bibitem{cvetyoum2} M. Cvetic and D. Youm, {\it Dyonic BPS saturated
black holes of heterotic string on a six-torus}, Phys. Rev. D{\bf 53}
(1996) 584 (hep-th/9507090); {\it BPS saturated and 
non-extreme states in abelian Kaluza-Klein theory and effective
$N=4$ supersymmetric string vacua}, (hep-th/9508058).

\bibitem{harrison} B. K. Harrison, J. Math. Phys. {\bf 9} (1968) 1744.

\bibitem{ross} S.F. Ross, {\it Pair production of black holes in a
$U(1)\otimes U(1)$ theory}, Phys. Rev. D{\bf 49} (1994) 6599
(hep-th/9401131).
 
\bibitem{yo} R. Emparan, {\it Composite black holes in external fields},
Nucl. Phys. {\bf B490} (1997) 365 (hep-th/9610170).

\bibitem{klop} R. Kallosh, A. Linde, T. Ort{\'\i}n, A. Peet
and A. Van Proeyen, {\it
Supersymmetry as a cosmic censor}, Phys. Rev. D{\bf 46} (1992) 5278 
(hep-th/9205027).

\bibitem{stelle} K. Stelle, {\it BPS branes in supergravity,} 
hep-th/9803116.

\end{thebibliography}
\end{document}